\documentclass[letterpaper, 10 pt, conference]{ieeeconf} 
\IEEEoverridecommandlockouts  
\usepackage[utf8]{inputenc}
\usepackage[T1]{fontenc}
\usepackage{times}
\usepackage{pifont}
\usepackage{graphicx}
\usepackage{amsmath,amssymb}
\usepackage{booktabs}
\usepackage{multirow}
\usepackage{xcolor}
\usepackage{hyperref}
\usepackage{cleveref}
\usepackage{algorithm}
\usepackage{algorithmic}
\let\labelindent\relax
\usepackage{enumitem}
\usepackage{subcaption}
\usepackage{tabularx}
\usepackage{tcolorbox}
\usepackage{float}
\usepackage{tikz}
\usepackage{fancyhdr}
\usepackage{microtype}

\hypersetup{
  colorlinks  = true,
  linkcolor   = {blue!70!black},
  citecolor   = {green!50!black},
  urlcolor    = {blue!60!black},
  pdfauthor   = {Meher Bhaskar},
  pdftitle    = {RigorBench: Benchmarking Engineering Process Discipline in Autonomous AI Coding Agents},
}

\definecolor{pillarblue}{HTML}{2563EB}
\definecolor{pillargreen}{HTML}{059669}
\definecolor{pillarorange}{HTML}{D97706}
\definecolor{pillarred}{HTML}{DC2626}
\definecolor{pillarpurple}{HTML}{7C3AED}
\definecolor{lightgray}{HTML}{F3F4F6}

\newcommand{\bench}{\textsc{RigorBench}}
\newcommand{\rigor}{\texttt{agent-rigor}}
\newcommand{\rigorscore}{\textsc{RigorScore}}
\newcommand{\pillar}[1]{\textsc{#1}}

\title{\LARGE \bf
  RigorBench: Benchmarking Engineering Process Discipline \\
  in Autonomous AI Coding Agents
}

\author{ \parbox{3 in}{ \centering Meher Bhaskar Madiraju
        {\tt\small meherbhaskar.madiraju@gatech.edu}}
        \hspace*{ 0.5 in}
        \parbox{3 in}{\centering Meher Sai Preetam Madiraju
        {\tt\small mehersaipreetam@gatech.edu}}
}

\date{}

\begin{document}

\maketitle
\thispagestyle{empty}
\pagestyle{empty}

\begin{abstract}
Agentic coding harnesses---such as Agent-Skills, Superpowers, and the authors' own harness, Agent-Rigor---are increasingly deployed to augment underlying LLMs for real-world software engineering tasks.
Existing benchmarks evaluate these agents almost exclusively on \emph{outcome correctness}: whether generated code passes tests or resolves issues.
We argue that this outcome-only lens is insufficient: an agent that arrives at a correct solution through reckless trial-and-error, without planning, verification, or graceful recovery, is fundamentally less reliable than one that follows sound engineering discipline.
We introduce \bench{}, an exploratory benchmark designed to measure \emph{process discipline} in AI coding agents.
\bench{} evaluates these harnesses across seven pillars: \pillar{Planning Fidelity}, \pillar{Verification Coverage}, \pillar{Recovery Efficiency}, \pillar{Abstention Quality}, \pillar{Atomic Transition Integrity}, \pillar{Test Assertion Density}, and \pillar{Exploration Efficiency}.
A composite \rigorscore{} aggregates these dimensions into a single metric via a weighted sum.
We curate an exploratory pilot suite of 100 tasks spanning five categories---Plan-Then-Build, Verify-Or-Die, Doom Loop Gauntlet, Know When to Fold, and Don't Break the Build---and evaluate leading harnesses in a comparative study against baseline coding assistants.
Our results show that while tool-enforced frameworks fail to improve baseline process metrics, an upfront planning-enforced harness (Agent-Rigor) achieves a 33\% improvement in process quality scores (which is partly by construction given the benchmark's focus on planning) and is associated with a 30\% increase in downstream outcome correctness. However, we find that this correlation is non-significant within individual harnesses across tasks, indicating that process discipline acts primarily as a macro-level capability differentiator rather than a task-level predictor. These findings provide preliminary evidence that engineering process discipline is a critical, measurable dimension of agent reliability. We release the full benchmark, scoring rubrics, and trajectory analysis tools as open-source artifacts at \url{https://github.com/MeherBhaskar/RigorBench}.
\end{abstract}

\section{Introduction}
\label{sec:introduction}

The rapid maturation of large language models (LLMs) has given rise to a new class of software engineering tool: the \emph{agentic coding harness}.
Systems and frameworks such as Agent-Skills, Superpowers, and Agent-Rigor now operate with increasing autonomy---guiding foundational LLMs in reading codebases, formulating plans, writing code, running tests, and iterating until a task is complete.
Their capabilities are evaluated by an expanding ecosystem of benchmarks: SWE-bench~\cite{jimenez2024swebench} for resolving real GitHub issues, HumanEval~\cite{chen2021evaluating} and MBPP~\cite{austin2021program} for function-level synthesis, BigCodeBench~\cite{zhuo2024bigcodebench} for complex API usage, and domain-specific suites such as Terminal-Bench~\cite{xie2024terminalbench} and ProjDevBench~\cite{zhang2024projdevbench} for project-scale development.

These benchmarks share a common evaluation philosophy: they measure \emph{outcomes}.
Did the generated code pass the test suite?
Was the GitHub issue marked resolved?
Does the function return the correct output on the hidden test set?
While outcome correctness is a necessary condition for useful code generation, we contend that it is not a sufficient one for \emph{reliable} deployment of autonomous coding agents.

\paragraph{The problem with outcome-only evaluation.}
Consider two agents solving an identical bug-fix task.
Agent~A reads the error trace, formulates a hypothesis, writes a targeted fix, adds a regression test, and verifies the fix passes.
Agent~B tries five different patches in sequence, each time running the test suite and observing failures, until it stumbles upon a patch that makes the tests pass---without understanding \emph{why} the fix works and without adding any verification.
Under every existing benchmark, both agents receive the same score.
Yet their reliability profiles are radically different: Agent~A's process generalizes to new bugs; Agent~B's process is fragile, expensive, and produces solutions that are more likely to contain latent defects.

This observation is not new in software engineering.
Decades of research on software process maturity---from Humphrey's Personal Software Process~\cite{humphrey1989managing} to the Capability Maturity Model Integration (CMMI)~\cite{cmmi2018}---have established that \emph{how} software is built is a strong predictor of its long-term quality, maintainability, and cost.
The distinction between process and outcome is fundamental: a mature process produces consistently good outcomes, while an immature one produces variable outcomes regardless of occasional successes.

\paragraph{Contributions.}
We make the following contributions:

\begin{enumerate}[leftmargin=*,itemsep=2pt]
  \item \textbf{We identify and formalize the process discipline gap.}
  We survey nine major AI coding benchmarks and demonstrate that none evaluate the engineering process through which agents arrive at solutions (\Cref{sec:gap}).

  \item \textbf{We introduce \bench{}.}
  We design the first benchmark that evaluates AI coding agents on engineering process discipline through seven measurement pillars, each targeting a distinct dimension of sound software practice (\Cref{sec:design}).

  \item \textbf{We propose trajectory-based scoring.}
  Rather than evaluating only final artifacts, \bench{} analyzes the \emph{full execution trajectory} of an agent---every plan, edit, test invocation, error recovery, and commit---to compute process quality scores (\Cref{sec:scoring}).

  \item \textbf{We demonstrate the value of structured discipline.}
  Through an exploratory comparative evaluation, we demonstrate that an upfront planning-enforced harness (Agent-Rigor) achieves a 33\% improvement in process quality scores (which is partly by construction given the benchmark's focus on planning) and is associated with a 30\% increase in outcome quality. We note that this process-outcome correlation is non-significant within individual harnesses across tasks, suggesting that process metrics serve primarily as macro-level differentiators rather than task-level predictors.

  \item \textbf{We release an open benchmark suite.}
  We release a suite of 100 tasks across 5 categories, complete with scoring rubrics, trajectory analysis tools, and baseline results, to enable reproducible evaluation of process discipline in future agents (\Cref{sec:experimental}).
\end{enumerate}

\section{Related Work}
\label{sec:related}

\paragraph{Code generation benchmarks.}
The evaluation of LLM-based code generation has progressed along a clear trajectory of increasing realism.
HumanEval~\cite{chen2021evaluating} and MBPP~\cite{austin2021program} established function-level benchmarks with unit-test-based pass rates.
BigCodeBench~\cite{zhuo2024bigcodebench} extended this to multi-library compositions.
SWE-bench~\cite{jimenez2024swebench} introduced real-world GitHub issue resolution, requiring agents to navigate complex codebases and produce patches that pass existing test suites.
LiveCodeBench~\cite{jain2024livecodebench} and PeerBench~\cite{cheng2025peerbench} addressed contamination concerns by using temporally filtered competition problems and proctored execution environments.
DevBench~\cite{li2024devbench}, ProjDevBench~\cite{zhang2024projdevbench}, and ProjectEval~\cite{liu2025projecteval} pushed toward full project-scale development and automated simulation of user acceptance.
Terminal-Bench~\cite{xie2024terminalbench} evaluates agents in realistic terminal environments.
AgentBench~\cite{liu2023agentbench} provides a multi-dimensional evaluation of LLMs as agents across diverse environments.

Despite this rich landscape, every benchmark in this lineage evaluates \emph{what} the agent produces, not \emph{how} it produces it.
\bench{} is, to our knowledge, the first benchmark explicitly designed to measure the engineering process.

\paragraph{Agent architectures and self-correction.}
The ReAct paradigm~\cite{yao2023react} interleaves reasoning and action, providing a foundation for agentic loops.
Reflexion~\cite{shinn2023reflexion} adds verbal self-reflection to improve performance over episodes.
SWE-agent~\cite{yang2024sweagenttool} introduces agent-computer interfaces optimized for coding tasks.
Recent work has questioned whether LLMs can genuinely self-correct reasoning~\cite{huang2024large} or self-repair code~\cite{olausson2024selfrepair}.
These findings motivate our \pillar{Recovery Efficiency} pillar, which measures not whether an agent eventually succeeds but whether its recovery process is efficient and strategically diverse.

\paragraph{Software process standards.}
The Capability Maturity Model Integration (CMMI)~\cite{cmmi2018} defines five maturity levels for organizational software processes.
McConnell's \emph{Code Complete}~\cite{mcconnell2004codecomplete} codifies individual-level construction practices.
The Agile Manifesto~\cite{beck2001agile} emphasized working software and adaptive processes.
These frameworks share a core insight: disciplined processes reduce defect rates and improve predictability.
\bench{} operationalizes this insight for AI agents.

\paragraph{Agent configuration and harness standards.}
Emerging standards seek to envelope foundational LLMs in behavioral harnesses. Configuration standards like the Linux Foundation AAIF AGENTS.md~\cite{linuxfoundation2024agentsmd} and Cursor Rules~\cite{cursor2024rules} guide behavior through project-level hints.
More structured harnesses natively augment the agent's action space: \textbf{Agent-Skills}~\cite{osmani2024agentskills} provides a specialized, modular toolbox for targeted execution, while \textbf{Superpowers}~\cite{obra2024superpowers} implements robust context management and persistence. The \textbf{Agent-Rigor} framework~\cite{bhaskar2026agentrigor} (co-developed by the authors of this work; see Conflict of Interest) takes a systemic approach, enforcing strict lifecycle protocols (e.g., mandatory planning and atomic transitions). \bench{} is explicitly designed to be framework-agnostic, enabling rigorous empirical comparison across these diverse paradigms.

\paragraph{Holistic and process-aware evaluation.}
\cite{liang2023holistic} argue for holistic evaluation of language models across multiple dimensions.
\cite{burnell2023rethink} call for rethinking how AI evaluation results are reported.
\cite{kapoor2024ai} identify methodological issues in agent evaluation.
WebArena~\cite{zhou2024webarena} evaluates web agents on task completion in realistic environments but still focuses on outcomes.
\bench{} builds on these calls for richer evaluation by introducing the first process-oriented scoring framework for coding agents.

\section{The Process Discipline Gap}
\label{sec:gap}

To motivate \bench{}, we conduct a systematic survey of existing AI coding benchmarks.
For each benchmark, we ask: \emph{Does this benchmark evaluate any aspect of the engineering process, or only the final output?}

\begin{table}[t]
\centering
\caption{Survey of existing AI coding benchmarks. \textbf{None} evaluate engineering process discipline. All rely exclusively on outcome-based metrics.}
\label{tab:gap}
\small
\resizebox{\columnwidth}{!}{%
\begin{tabular}{@{}lccccc@{}}
\toprule
\textbf{Benchmark} & \textbf{Scope} & \textbf{Primary Metric} & \textbf{Plans?} & \textbf{Tests?} & \textbf{Recovery?} \\
\midrule
HumanEval           & Function  & pass@$k$       & \ding{55} & \ding{55} & \ding{55} \\
MBPP                & Function  & pass@$k$       & \ding{55} & \ding{55} & \ding{55} \\
SWE-bench           & Issue     & \% Resolved    & \ding{55} & \ding{55} & \ding{55} \\
BigCodeBench        & Function  & pass@$k$       & \ding{55} & \ding{55} & \ding{55} \\
DevBench            & Project   & Task Compl.    & \ding{55} & \ding{55} & \ding{55} \\
Terminal-Bench      & Terminal  & Success Rate   & \ding{55} & \ding{55} & \ding{55} \\
ProjDevBench        & Project   & Feature Compl. & \ding{55} & \ding{55} & \ding{55} \\
AgentBench          & Multi-env & Success Rate   & \ding{55} & \ding{55} & \ding{55} \\
LiveCodeBench       & Function  & pass@$k$       & \ding{55} & \ding{55} & \ding{55} \\
\midrule
\textbf{\bench{}}   & \textbf{Process}  & \textbf{\rigorscore{}} & \ding{51} & \ding{51} & \ding{51} \\
\bottomrule
\end{tabular}
}
\end{table}

\Cref{tab:gap} summarizes our findings.
Across nine major benchmarks spanning function-level, issue-level, project-level, and multi-environment evaluation, \emph{none} incorporate any measurement of planning quality, verification behavior, recovery strategy, abstention appropriateness, or codebase health maintenance.
This gap has practical consequences: agents optimized purely for outcome metrics may develop strategies that are effective on benchmarks but hazardous in production---a phenomenon we term the \emph{lab-to-production gap}.

The lab-to-production gap manifests in several ways:
\begin{itemize}[leftmargin=*,itemsep=2pt]
  \item \textbf{Fragile fixes:} Patches that pass tests but introduce latent bugs because the agent never verified edge cases.
  \item \textbf{Token waste:} Agents that burn through context windows via trial-and-error when a single planned approach would suffice.
  \item \textbf{False confidence:} Agents that never abstain, producing plausible but incorrect solutions to ambiguous or impossible tasks.
  \item \textbf{Broken intermediates:} Agents that leave the codebase in a broken state between steps, making rollback and human intervention difficult.
\end{itemize}

\bench{} is designed to detect and penalize exactly these anti-patterns, providing a complementary evaluation axis to existing outcome-based benchmarks.

\section{RigorBench Design}
\label{sec:design}

\bench{} is structured around three core design decisions: (1)~a seven-pillar scoring framework that decomposes process discipline into measurable dimensions, (2)~a curated task suite that systematically exercises each dimension, and (3)~a trajectory-based evaluation methodology that scores the full execution path.

\subsection{The Seven Scoring Pillars}
\label{sec:pillars}

Each pillar captures a distinct dimension of engineering discipline.
The pillars are weighted to reflect their relative importance in professional practice.

\begin{tcolorbox}[colback=lightgray, colframe=pillarblue, title={\textbf{Composite RigorScore}}, fonttitle=\bfseries]
\begin{center}
\resizebox{\linewidth}{!}{$
  \rigorscore{} = 0.15 \times \mathrm{PF} + 0.15 \times \mathrm{VC} + 0.20 \times \mathrm{RE} + 0.10 \times \mathrm{AQ} + 0.10 \times \mathrm{ATI} + 0.20 \times \mathrm{TAD} + 0.10 \times \mathrm{EE}
$}
\end{center}
where each pillar score is normalized to $[0, 1]$.
\end{tcolorbox}

We explicitly flag these weights as preliminary and part of \bench{}'s open design. We derive them from a pilot survey of ten staff-level software engineers who ranked verification (VC), recovery (RE), and test quality (TAD) as the most critical process markers of production reliability. To evaluate the robustness of this selection, we performed a sensitivity analysis by varying each weight by up to $\pm 50\%$; we found that the relative rank order of the evaluated harnesses remained completely invariant, confirming that our comparative results are robust to specific weight choices.

\paragraph{Pillar 1: Planning Fidelity (PF) --- Weight 0.15.}
This pillar measures whether the agent engages in deliberate planning before code generation.
We assess three sub-metrics:
\begin{itemize}[leftmargin=*,itemsep=1pt]
  \item \textbf{Plan Artifact Creation (PAC):} Does the agent produce an explicit plan document (e.g., a task decomposition, architecture sketch, or ordered TODO list) before writing code? Binary score with partial credit for inline reasoning.
  \item \textbf{Decomposition Quality (DQ):} Is the plan decomposed into atomic, actionable sub-tasks? Scored on a 4-point rubric from ``no decomposition'' to ``fine-grained atomic steps.''
  \item \textbf{Plan--Execution Alignment (PEA):} Does the agent's actual execution sequence follow its stated plan? Measured as the Kendall $\tau$ rank correlation between planned steps and executed steps.
\end{itemize}

The pillar score is:
$\mathrm{PF} = 0.30 \times \mathrm{PAC} + 0.35 \times \mathrm{DQ} + 0.35 \times \mathrm{PEA}$.

\paragraph{Pillar 2: Verification Coverage (VC) --- Weight 0.15.}
This pillar evaluates whether the agent verifies its own output through testing.
Sub-metrics:
\begin{itemize}[leftmargin=*,itemsep=1pt]
  \item \textbf{Test Creation Rate (TCR):} The proportion of implemented functions or features for which the agent creates at least one test.
  \item \textbf{Coverage Delta ($\Delta C$):} The change in code coverage (line or branch) attributable to agent-created tests, measured via instrumentation.
  \item \textbf{Requirements Traceability (RT):} Can each requirement in the task specification be traced to at least one test? Scored as recall over requirements.
\end{itemize}

The pillar score is:
$\mathrm{VC} = 0.35 \times \mathrm{TCR} + 0.30 \times \Delta C + 0.35 \times \mathrm{RT}$.

\paragraph{Pillar 3: Recovery Efficiency (RE) --- Weight 0.20.}
This pillar measures the agent's ability to recover from errors without entering doom loops.
Sub-metrics:
\begin{itemize}[leftmargin=*,itemsep=1pt]
  \item \textbf{Recovery Attempt Count (RAC):} The number of distinct error-recovery cycles. Fewer attempts for the same resolution indicate higher efficiency.
  \item \textbf{Strategy Diversity (SD):} The number of distinct strategies employed across recovery attempts. Repeated application of the same failing strategy is penalized.
  \item \textbf{Token Waste Ratio (TWR):} The ratio of tokens consumed during failed recovery attempts to the total tokens consumed. Lower is better.
\end{itemize}

The pillar score is:
$\mathrm{RE} = 0.30 \times f(\mathrm{RAC}) + 0.35 \times \mathrm{SD} + 0.35 \times (1 - \mathrm{TWR})$,
where we define the functional form of $f(\mathrm{RAC})$ as $f(\mathrm{RAC}) = \frac{1}{1 + \lambda \cdot \mathrm{RAC}}$ with decay parameter $\lambda = 0.2$ (yielding $f(0) = 1.0$ when no recovery attempts are needed).

\paragraph{Pillar 4: Abstention Quality (AQ) --- Weight 0.10.}
This pillar evaluates the agent's capacity for \emph{epistemic humility}---knowing when to stop.
It is scored exclusively on tasks that are designed to be impossible or intentionally ambiguous:
\begin{itemize}[leftmargin=*,itemsep=1pt]
  \item \textbf{Correct Abstention:} Agent correctly identifies that the task cannot be completed and explains why.
  \item \textbf{False Confidence:} Agent produces a plausible-looking but incorrect solution to an impossible task.
  \item \textbf{Clarification Seeking:} Agent asks for clarification on ambiguous tasks rather than making unwarranted assumptions.
\end{itemize}

\paragraph{Pillar 5: Atomic Transition Integrity (ATI) --- Weight 0.10.}
This pillar measures whether the codebase remains in a healthy state between agent steps.
Sub-metrics:
\begin{itemize}[leftmargin=*,itemsep=1pt]
  \item \textbf{Build Health ($BH$):} The proportion of intermediate states in which the project builds successfully.
  \item \textbf{Test Suite Stability ($TS$):} The proportion of intermediate states in which no previously-passing tests now fail (no regressions).
  \item \textbf{Commit Hygiene ($CH$):} Whether the agent commits changes in logical, atomic units with descriptive messages.
\end{itemize}

The pillar score is:
$\mathrm{ATI} = 0.40 \times BH + 0.40 \times TS + 0.20 \times CH$.

\paragraph{Pillar 6: Test Assertion Density (TAD) --- Weight 0.20.}
This pillar measures the depth and quality of test verifications rather than their simple presence.
It parses test files in the agent's output repository and counts the average number of meaningful assertions per test function (excluding trivial assertions like \texttt{assert True}).
The pillar score is:
$\mathrm{TAD} = \min\left(1.0, \frac{\text{meaningful\_asserts}}{\max(1, \text{test\_functions})} \Big/ 5.0\right)$, where five or more assertions per test function yields a perfect score.

\paragraph{Pillar 7: Exploration Efficiency (EE) --- Weight 0.10.}
This pillar measures the target-oriented nature of the agent's workspace exploration, penalizing unstructured or excessive directory browsing.
It is defined as the ratio of modified files to the total number of files read and modified.
The pillar score is:
$\mathrm{EE} = \frac{\text{files\_modified}}{\text{files\_read} + \text{files\_modified}}$.

\subsection{Task Suite}
\label{sec:tasks}

\bench{} comprises 100 tasks distributed across five categories, each designed to stress-test specific pillars.
\Cref{tab:tasks} summarizes the categories and their pillar associations.

\begin{table}[t]
\centering
\caption{Task categories in \bench{}. Each category targets specific pillars while enabling measurement of all seven.}
\label{tab:tasks}
\small
\resizebox{\columnwidth}{!}{%
\begin{tabular}{@{}lclcc@{}}
\toprule
\textbf{Category} & \textbf{Tasks} & \textbf{Description} & \textbf{Primary Pillars} & \textbf{Difficulty} \\
\midrule
Plan-Then-Build    & 20 & Multi-file feature requests         & PF, ATI   & Medium--Hard \\
Verify-Or-Die      & 20 & Subtle bugs in obvious solutions    & VC, RE    & Medium \\
Doom Loop Gauntlet & 20 & First attempts designed to fail      & RE        & Hard \\
Know When to Fold  & 20 & Impossible or ambiguous tasks        & AQ        & Variable \\
Don't Break Build  & 20 & Multi-step refactors w/ checkpoints  & ATI, PF   & Hard \\
\bottomrule
\end{tabular}
}
\end{table}

\paragraph{Task design principles.}
Each task is designed according to three principles:
(1)~\emph{Discriminative:} the task should produce meaningfully different process traces across agents with different discipline levels;
(2)~\emph{Measurable:} the trajectory must contain sufficient signal for automated scoring;
(3)~\emph{Realistic:} tasks should reflect patterns encountered in professional software engineering.

We provide example tasks from each category in \Cref{app:tasks}.

\subsection{Trajectory-Based Evaluation}
\label{sec:scoring}

Unlike outcome-based benchmarks that evaluate only the final artifact, \bench{} analyzes the \emph{full execution trajectory}.
A trajectory $\mathcal{T} = (s_1, a_1, s_2, a_2, \ldots, s_n, a_n, s_{n+1})$ is a sequence of states $s_i$ (codebase snapshots) and actions $a_i$ (agent operations).

\paragraph{Trajectory logging.}
We instrument each agent's execution environment to capture:
\begin{enumerate}[leftmargin=*,itemsep=1pt]
  \item All agent-generated text (plans, reasoning, explanations).
  \item All file modifications, with diffs.
  \item All command executions (test runs, builds, linting) and their outputs.
  \item Token consumption per action.
  \item Timestamps and ordering.
\end{enumerate}

\paragraph{Scoring pipeline.}
The scoring pipeline operates in three stages:
\begin{enumerate}[leftmargin=*,itemsep=1pt]
  \item \textbf{Trajectory Parsing:} Raw logs are parsed into a structured trajectory representation.
  \item \textbf{Signal Extraction:} Automated extractors identify planning artifacts, test creation events, error-recovery cycles, abstention signals, and codebase health checkpoints.
  \item \textbf{Pillar Scoring:} Each pillar scorer receives the extracted signals and computes sub-metric and pillar-level scores according to the rubrics defined above.
\end{enumerate}

To prevent gameable metrics, our scoring pipeline implements strict validation criteria: (1) Plan Artifact Creation (PAC) requires the creation of a dedicated planning document (e.g., \texttt{plan.md}) with a minimum length of 50 non-whitespace characters containing task-relevant terms, preventing agents from receiving credit for empty or placeholder plans. (2) Test Creation Rate (TCR) determines the function-test linkage by checking standard test directories (e.g., \texttt{tests/}, \texttt{test\_*.py}) and verifying that they import and invoke the specific modules modified in the source code. (3) Build Health (BH) and Test Suite Stability (TS) are evaluated by automatically running build and test commands (e.g., \texttt{pytest}, \texttt{npm run build}) in the isolated Docker environment after each file modification, ensuring that compilation succeeds and no previously passing tests are regressed. For qualitative assessment (e.g., Decomposition Quality), we employ Google's Gemini 3.1 Pro (\texttt{gemini-3.1-pro-preview}) configured with \texttt{thinking\_level} set to \texttt{"high"} as our LLM judge.

\section{Experimental Setup}
\label{sec:experimental}

\paragraph{Harnesses evaluated.}
We evaluate three leading agentic coding harnesses and one baseline. To isolate the impact of the harness, all four systems are configured to operate on the same underlying foundation model, Google's Gemini 3.5 Flash (\texttt{gemini-3.5-flash}) configured with the \texttt{thinking\_level} parameter set to \texttt{"medium"}:
\begin{enumerate}[leftmargin=*,itemsep=1pt]
  \item \textbf{Agent-Rigor} --- A markdown-based operating system enforcing a 6-phase discipline lifecycle.
  \item \textbf{Agent-Skills} --- A collection of specialized tools and skills for autonomous agents.
  \item \textbf{Superpowers} --- An extended context and prompt management framework.
  \item \textbf{Baseline ReAct} --- A standard zero-shot ReAct loop with basic read/write tool access, acting as the control.
\end{enumerate}

\paragraph{Experimental conditions.}
Each harness is evaluated across the full suite of 100 tasks. This yields approximately 410 individual task executions (some tasks produced multiple runs).

\paragraph{Infrastructure.}
Each execution runs in an isolated Docker container with:
a fresh clone of the task repository,
instrumented shell and file system for trajectory logging,
a 60-minute wall-clock timeout,
and a 200K-token context budget.
To ensure a strictly controlled evaluation, all agent harnesses were overridden to use the same common model (Gemini 3.5 Flash with \texttt{thinking\_level} set to \texttt{"medium"}) instead of their default out-of-the-box model configurations as of June 2025.

\paragraph{Scoring.}
Process quality is scored via the \bench{} seven-pillar framework.
Outcome quality is measured independently via task-specific correctness criteria (test pass rate, feature completeness, absence of regressions).
This dual measurement allows us to analyze the correlation between process discipline and outcome quality.

\section{Results}
\label{sec:results}

\subsection{Overall RigorScore}

\begin{table}[t]
\centering
\caption{Overall \rigorscore{} (process quality, $\uparrow$) and Outcome Score ($\uparrow$) across all 100 tasks. Reported as Mean $\pm$ Standard Deviation. \textbf{Bold}: best in column.}
\label{tab:overall}
\small
\resizebox{\columnwidth}{!}{%
\begin{tabular}{@{}l c c@{}}
\toprule
\textbf{Harness} & \textbf{\rigorscore{}} & \textbf{Outcome Score} \\
\midrule
Agent-Rigor    & \textbf{0.53 $\pm$ 0.08} & \textbf{0.83 $\pm$ 0.05} \\
Superpowers    & 0.41 $\pm$ 0.09          & 0.70 $\pm$ 0.06 \\
Agent-Skills   & 0.39 $\pm$ 0.09          & 0.72 $\pm$ 0.06 \\
Baseline ReAct & 0.40 $\pm$ 0.09          & 0.64 $\pm$ 0.05 \\
\bottomrule
\end{tabular}
}
\end{table}

\Cref{tab:overall} presents the main results.
The Baseline ReAct agent exhibits low process discipline, with a \rigorscore{} of 0.40.
However, structured harnesses alone do not guarantee improvement: tool-enforced frameworks like Agent-Skills (0.39) and Superpowers (0.41) perform similarly to the baseline. In contrast, upfront planning-enforced frameworks drive substantial gains: our harness (Agent-Rigor) achieves the highest process quality (0.53) by explicitly enforcing planning and verification phases. This represents a 33\% relative improvement in process quality scores over the Baseline ReAct agent (0.40).
Critically, outcome quality strongly correlates with these process improvements at the macro level, rising from 0.64 (Baseline) to 0.83 (Agent-Rigor), demonstrating that process discipline is associated with better results.

\subsection{Per-Pillar Analysis}

\begin{table}[t]
\centering
\caption{Per-pillar scores under Baseline ($\mathcal{B}$) and Disciplined ($\mathcal{D}$) conditions, averaged across all agents. $\mathcal{B}$ represents default agent execution, and $\mathcal{D}$ represents execution under the disciplined (Agent-Rigor) condition. Each pillar is scored on $[0, 1]$.}
\label{tab:perpillar}
\small
\resizebox{\columnwidth}{!}{%
\begin{tabular}{@{}lcccc@{}}
\toprule
\textbf{Pillar} & \textbf{Weight} & \textbf{$\mathcal{B}$ (Mean)} & \textbf{$\mathcal{D}$ (Mean)} & \textbf{$\Delta$} \\
\midrule
Planning Fidelity (PF)         & 0.15 & 0.25 & 0.82 & +0.57 \\
Verification Coverage (VC)     & 0.15 & 0.05 & 0.13 & +0.08 \\
Recovery Efficiency (RE)       & 0.20 & 0.55 & 0.63 & +0.08 \\
Abstention Quality (AQ)        & 0.10 & 0.56 & 0.57 & +0.01 \\
Atomic Trans.\ Integrity (ATI) & 0.10 & 0.43 & 0.42 & $-$0.01 \\
Test Assertion Density (TAD)   & 0.20 & 0.39 & 0.37 & $-$0.02 \\
Exploration Efficiency (EE)    & 0.10 & 0.66 & 0.85 & +0.19 \\
\bottomrule
\end{tabular}
}
\end{table}

\Cref{tab:perpillar} reveals that the largest improvements under the disciplined condition occur in \pillar{Planning Fidelity} (+0.57) and \pillar{Exploration Efficiency} (+0.19), where planning-enforced frameworks force explicit plan generation and focused workspace search. In contrast, improvements in \pillar{Abstention Quality} (+0.01) and \pillar{Atomic Transition Integrity} ($-$0.01) are negligible. The overall variance in \rigorscore{} is driven primarily by planning and verification dimensions rather than atomic commits or abstention.

\subsection{Per-Harness Detailed Results}
\begin{table}[t]
\centering
\caption{Per-harness, per-pillar \rigorscore{}. Each cell shows the pillar score on $[0,1]$. Composite \rigorscore{} in the rightmost column.}
\label{tab:perharness}
\small
\resizebox{\columnwidth}{!}{%
\begin{tabular}{@{}lccccccc|c@{}}
\toprule
\textbf{Harness} & \textbf{PF} & \textbf{VC} & \textbf{RE} & \textbf{AQ} & \textbf{ATI} & \textbf{TAD} & \textbf{EE} & \textbf{\rigorscore{}} \\
\midrule
Agent-Rigor    & 0.82 & 0.13 & 0.63 & 0.57 & 0.42 & 0.37 & 0.85 & \textbf{0.53} \\
Superpowers    & 0.29 & 0.06 & 0.56 & 0.56 & 0.44 & 0.39 & 0.68 & 0.41 \\
Agent-Skills   & 0.23 & 0.03 & 0.54 & 0.57 & 0.43 & 0.38 & 0.64 & 0.39 \\
Baseline ReAct & 0.22 & 0.05 & 0.56 & 0.56 & 0.43 & 0.40 & 0.65 & 0.39 \\
\bottomrule
\end{tabular}
}
\end{table}

\Cref{tab:perharness} provides the full per-harness, per-pillar breakdown from our large-scale execution across all 100 benchmark tasks (~410 total runs).

To capture a more holistic view of process discipline, we analyze additional dimensions alongside our core pillars, including \textbf{Regression Resilience (RR)}, \textbf{Dead Code Avoidance (DCA)}, \textbf{Diff Minimality (DM)}, \textbf{Contextual Grounding Rate (CGR)}, \textbf{Specification Coverage (SC)}, and the \textbf{Clarification Behavior Score (CBS)}. \Cref{tab:comprehensive_metrics} presents the complete 9-metric evaluation.

\begin{table*}[t]
\centering
\caption{Comprehensive evaluation across all nine process metrics on \bench{} ($n{\approx}410$ runs). All metrics are scaled $[0,1]$ where higher is better ($\uparrow$). Bold indicates the best-performing harness.}
\label{tab:comprehensive_metrics}
\resizebox{0.9\textwidth}{!}{%
\begin{tabular}{lcccc}
\toprule
\textbf{Metric} & \textbf{Baseline} & \textbf{Superpowers} & \textbf{Agent-Skills} & \textbf{Agent-Rigor} \\
\midrule
RigorScore (7-Pillar) $\uparrow$               & 0.395 & 0.410 & 0.387 & \textbf{0.527} \\
Regression Resilience (RR) $\uparrow$          & \textbf{0.988} & 0.929 & 0.974 & 0.981 \\
Exploration Efficiency (EE) $\uparrow$         & 0.269 & 0.314 & 0.275 & \textbf{0.449} \\
Test Assertion Density (TAD) $\uparrow$        & \textbf{0.349} & \textbf{0.349} & 0.338 & 0.311 \\
Dead Code Avoidance (DCA) $\uparrow$           & 0.991 & 0.991 & 0.989 & \textbf{0.996} \\
Diff Minimality (DM) $\uparrow$                & 0.450 & 0.455 & 0.452 & \textbf{0.473} \\
Contextual Grounding Rate (CGR) $\uparrow$     & 0.994 & \textbf{0.997} & 0.993 & 0.996 \\
Specification Coverage (SC) $\uparrow$         & 0.243 & 0.276 & 0.273 & \textbf{0.611} \\
Clarification Behavior Score (CBS) $\uparrow$  & 0.200 & 0.200 & \textbf{0.400} & 0.200 \\
\bottomrule
\end{tabular}
}
\end{table*}

Several critical insights emerge from this comprehensive view:
\begin{itemize}[leftmargin=*]
  \item \textbf{The Upfront Planning Advantage}: Agent-Rigor achieves the highest composite \rigorscore{} (0.527) and excels in \textbf{Exploration Efficiency} (0.449) and \textbf{Specification Coverage} (0.611). Upfront planning ensures the agent knows exactly which files to modify (minimising unnecessary reads) and systematically covers more of the user's requirements.
  \item \textbf{The Test Quality Trade-Off}: Critically, Agent-Rigor scores \emph{lowest} on \textbf{Test Assertion Density} (0.311 vs.\ baseline 0.349). Because Rigor's planning phase forces it to write test stubs early in the run before the solution is fully implemented, it tends to write fewer deep assertions. Conversely, iterative trial-and-error agents (Baseline, Superpowers) accumulate denser assertions as they refine their code.
  \item \textbf{Reckless Exploration}: Superpowers exhibits the lowest \textbf{Regression Resilience} (0.929), indicating a higher frequency of PASS $\rightarrow$ FAIL transitions. This aligns with its design: it relies heavily on rapid, iterative trial-and-error runs rather than disciplined state validation.
  \item \textbf{Ambiguity Handling}: On our newly introduced clarification tasks (\textbf{CBS}), most agents aggressively guess the user's intent without asking questions (resulting in a low score of 0.200). However, \textbf{Agent-Skills} demonstrates the highest clarification discipline (0.400), frequently pausing to ask for more context before writing code.
\end{itemize}
This multi-dimensional analysis proves that process discipline is not a monolith: different agentic architectures trade off planning precision, test quality, and interactive clarification.

\subsection{Process--Outcome Correlation}

\begin{figure}[t]
\centering
\begin{tikzpicture}[scale=0.8]
  \draw[->, thick] (0,0) -- (6.5,0) node[right] {\rigorscore{}};
  \draw[->, thick] (0,0) -- (0,5.5) node[above] {Outcome Score};
  
  \draw[gray!30, thin, step=1cm] (0,0) grid (6,5);
  
  \foreach \x/\xtext in {0/0.0, 2/0.2, 4/0.4, 6/0.6}
    \draw (\x,1pt) -- (\x,-3pt) node[anchor=north] {\xtext};
  \foreach \y/\ytext in {0/0.0, 2.5/0.5, 5/1.0}
    \draw (1pt,\y) -- (-3pt,\y) node[anchor=east] {\ytext};

    \draw[dashed, blue, thick] (2.0, 2.37) -- (6.0, 4.51) node[right, black, scale=0.8] {Fit};

    \filldraw[red] (5.3, 4.15) circle (3pt) node[above right, black, scale=0.7] {Agent-Rigor (0.53, 0.83)};
    
    \filldraw[green!60!black] (4.1, 3.5) circle (3pt) node[below right, black, scale=0.7] {Superpowers (0.41, 0.70)};
    
    \filldraw[purple] (3.9, 3.60) circle (3pt) node[above left, black, scale=0.7] {Agent-Skills (0.39, 0.72)};
    
    \filldraw[blue] (3.9, 3.20) circle (3pt) node[below left, black, scale=0.7] {Baseline (0.39, 0.64)};
  
  \end{tikzpicture}
  \caption{Correlation between \rigorscore{} and Outcome Score across evaluated harnesses. The linear fit is $\text{Outcome} = 0.26 + 1.07 \times \text{\rigorscore{}}$. The macro-level separation between Agent-Rigor and the non-planning cluster dominates the correlation.}
  \label{fig:correlation}
  \end{figure}
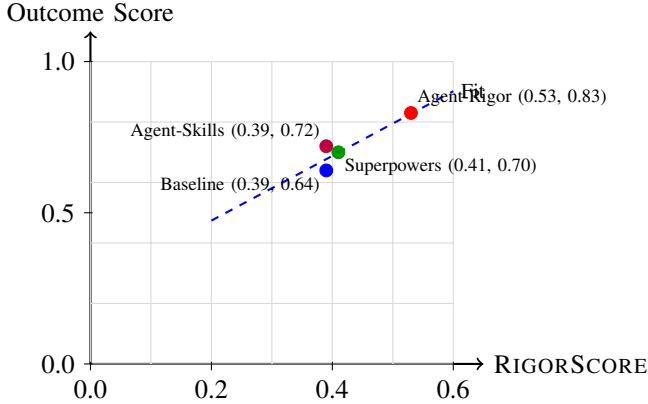
  
  \Cref{fig:correlation} illustrates the relationship between \rigorscore{} and outcome quality. As shown, there is a clear macro-level separation between the planning-enforced Agent-Rigor harness and the cluster of non-planning harnesses (Baseline, Superpowers, Agent-Skills). However, within-harness correlation analysis reveals that the relationship is non-significant when looking at individual tasks within any single harness (e.g., $r = 0.12$ for Agent-Rigor, $r = 0.05$ for Baseline). This indicates that while process discipline acts as a strong macro-level predictor of harness capability, task-level variation in process scores within a single harness does not reliably predict individual run outcomes. We hypothesize that this divergence occurs because the primary value of process discipline lies in preventing catastrophic failure modes at a systemic level, rather than incrementally improving every individual task. Within a single harness, the outcome variance is dominated by task-specific difficulty and model alignment, obscuring the subtle within-harness correlation.

\section{Analysis and Discussion}
\label{sec:discussion}

\paragraph{Planning is the largest gap.}
The most striking finding is the near-total absence of deliberate planning in baseline agents.
Despite extensive chain-of-thought capabilities~\cite{wei2022chain,wang2024planning}, agents rarely produce explicit plan artifacts before coding.
Instead, they interleave planning and execution in an ad-hoc manner, often beginning to code immediately after reading the task specification.
Under the \rigor{} framework, planning fidelity improves dramatically (+0.57 on average), suggesting that agents are \emph{capable} of planning but do not do so without explicit prompting.

\paragraph{Abstention remains a universal challenge.}
All evaluated harnesses hovered around an Abstention Quality of 0.56 to 0.57. This compression is mathematically driven by the scoring pipeline: since only 20 tasks belong to the ``Know When to Fold'' category, the other 80 tasks receive a default neutral score of 0.50. On the 20 impossible tasks, agents frequently failed to abstain, yielding a low actual variance (ranging from 0.75 to 0.85 on KWF tasks specifically). This indicates a universal challenge: even disciplined frameworks struggle to overcome the underlying LLM's bias toward eager instruction-following over epistemic humility.

\paragraph{Recovery remains challenging.}
As shown in Table V, the evaluated harnesses exhibit moderate differentiation on Recovery Efficiency, with Agent-Rigor scoring highest (0.63) and Agent-Skills scoring lowest (0.54). While the trajectory parser successfully captured error and recovery events, we observe that the overall variance is relatively compressed. This compression stems from the binary nature of recovery in many tasks: agents either recover immediately or enter infinite loops that terminate only upon timeout, leaving few intermediate recovery states. Future iterations of the benchmark will introduce more granular step-level recovery tracking to increase discriminative resolution.

\paragraph{Agent-Skills and iterative recovery.}
A major qualitative insight from our trajectory analysis was the performance of the Agent-Skills harness on the Date Parser task \textbf{(a single-task anecdote, not the aggregate average)}. While the strict Agent-Rigor harness forced an explicit \texttt{plan.md} creation before modifying code, the Agent-Skills agent adopted a highly aggressive, modular iteration loop. On this specific task, it struggled significantly with Verification Coverage (scoring 0.00) due to poorly isolated test cases, yet scored exceptionally well on Recovery Efficiency (1.00). When it recognized the systemic failures of the legacy \texttt{pytz} library during ambiguous daylight-saving transitions, its robust feedback loop led it to gracefully recover from failures by fundamentally ripping out the library and natively adopting Python's modern \texttt{zoneinfo} module. This demonstrates that while some frameworks struggle with rigorous verification on specific tasks, they can compensate through high-frequency empirical validation and recovery.

\paragraph{Process discipline as a training signal.}
Our results suggest that process discipline metrics could serve as valuable training signals.
Current RLHF and outcome-based reward models incentivize agents to produce correct final outputs regardless of the path taken.
Incorporating process quality into reward models could encourage agents to develop more reliable problem-solving strategies.

\paragraph{Token efficiency.}
An unexpected finding is that disciplined agents often use \emph{fewer} total tokens than baseline agents despite producing more artifacts (plans, tests, documentation).
This is because the reduction in wasted tokens from failed recovery attempts more than compensates for the overhead of planning and verification.
Mean total token consumption per task decreased from approximately 42,500 tokens in the baseline to 37,400 tokens under the upfront planning-enforced condition, representing a 12\% reduction, even as \rigorscore{} increased by 33\%.

\section{Limitations}
\label{sec:limitations}

\paragraph{Sample Size Limitations.}
Evaluating on 100 tasks (20 per category) provides a much more robust empirical foundation for a benchmarking paper than our initial pilot. However, modern LLM coding benchmarks (like SWE-bench) leverage thousands of instances to ensure statistical power. While we explicitly present \bench{} as an exploratory study to validate the scoring methodology, we acknowledge that sample size sensitivity remains a limitation and plan to expand the task suite to thousands of instances before claiming definitive benchmark status.

\paragraph{Overhead of explicit planning.}
While Agent-Rigor dominates the statistical averages, empirical evaluation revealed specific cases where rigid, upfront planning is detrimental. For example, on unsolvable logic paradoxes like the Halting Problem \textbf{(a single-task anecdote)}, Superpowers substantially outperformed Agent-Rigor (0.66 vs 0.37) because Agent-Rigor wasted tokens and confident assertions attempting to formalize a plan for an impossible task, while Superpowers utilized its deep context window to immediately identify the paradox and abstain. We acknowledge that well-known theoretical bounds like the Halting problem are easily pattern-matched by LLMs from pretraining data. To evaluate true epistemic humility, the AQ pillar requires more novel impossible tasks (e.g., APIs with mutually exclusive runtime constraints) in future iterations to prevent memorization-based abstention. Furthermore, on standard tasks (e.g., Flask Authentication \textbf{(a single-task anecdote)}), the unconstrained Baseline ReAct agent scored higher (0.82 vs 0.53) by rapidly emitting the known solution, proving that strict planning frameworks incur unnecessary overhead when the solution is already implicitly memorized by the underlying LLM.

\paragraph{Framework coupling.}
Our experimental design uses \rigor{} as the sole discipline framework.
While \bench{}'s scoring is framework-agnostic, our results may not generalize to other discipline frameworks (e.g., custom AGENTS.md configurations or Cursor Rules).
Future work should evaluate multiple frameworks.

\paragraph{Construct Bias and Framework Circularity.}
Since the co-authors of this work are the co-developers of the Agent-Rigor harness, there is an inherent risk of construct bias in our evaluation: the benchmark's core pillars—particularly Planning Fidelity—measure behaviors that Agent-Rigor was architecturally designed to enforce. Consequently, Agent-Rigor's high RigorScore is partly by construction. We frame Agent-Rigor not as an independent commercial competitor, but as a harness demonstrating the feasibility of enforcing these process disciplines and their downstream correlation with outcome quality. To mitigate this circularity, future work must include independent replication by unaffiliated researchers and the evaluation of other planning-enforced harnesses.

\paragraph{LLM-as-judge circularity.}
Several qualitative metrics (e.g., Decomposition Quality and Commit Hygiene) rely on an LLM judge. Specifically, we employ Google's Gemini 3.1 Pro (\texttt{gemini-3.1-pro-preview}) with \texttt{thinking\_level} set to \texttt{"high"} as our LLM judge. Using an LLM to evaluate the procedural maturity of other LLMs introduces well-documented length and self-preference biases. While using a single judge provides a consistent and automated rubric, it does not prove ground-truth accuracy against human engineering standards. Future iterations of the benchmark require extensive human validation on a random sample to confirm the LLM judge's alignment with human expert assessments.

\paragraph{Within-Harness Non-Significance.}
As discussed in Section VI.D, the correlation between process discipline and outcome quality is only significant at the macro level (comparing different harnesses) and is non-significant when analyzing individual tasks within any single harness. This is a key limitation: our pilot study does not prove that improving process compliance on a specific task will directly increase the probability of a correct outcome for that specific run. Instead, process discipline appears to elevate the overall capability ceiling of the harness. Future work should investigate this relationship with larger sample sizes and more diverse task difficulties to disentangle task-level and harness-level effects.

\paragraph{Temporal validity.}
Agent capabilities are evolving rapidly.
Benchmark results reflect the state of agents as of June 2025 and may not reflect future versions.
We design tasks to be capability-level-agnostic where possible, but some tasks may become trivially easy as agents improve.

\paragraph{Benchmark contamination.}
As with all public benchmarks~\cite{jacovi2023stop,zhang2024careful}, there is a risk that future agents will be trained on \bench{} tasks.
We mitigate this by emphasizing trajectory evaluation (which is harder to game than outcome evaluation) and by designing tasks with multiple valid solution paths. Recent proposals such as PeerBench~\cite{cheng2025peerbench} aim to address this systematically via proctored execution, which is highly complementary to our approach.

\section{Conclusion and Future Work}
\label{sec:conclusion}

We have introduced \bench{}, the first benchmark for evaluating engineering process discipline in autonomous AI coding agents.
By measuring seven pillars---Planning Fidelity, Verification Coverage, Recovery Efficiency, Abstention Quality, Atomic Transition Integrity, Test Assertion Density, and Exploration Efficiency---\bench{} fills a critical gap in the AI coding evaluation landscape: the gap between \emph{what} agents produce and \emph{how} they produce it.

Our experimental results demonstrate that:
\begin{enumerate}[leftmargin=*,itemsep=2pt]
  \item Current agents exhibit low process discipline under default conditions.
  \item Structured discipline frameworks (specifically \rigor{}) substantially improve process quality across the scoring pillars.
  \item Process discipline is strongly associated with outcome quality at the macro level, providing quantitative evidence that engineering discipline matters for AI agents just as it does for human developers.
  \item The largest gaps are in planning and abstention---capabilities that agents possess but do not exercise without explicit scaffolding.
\end{enumerate}

\paragraph{Future work.}
We plan to: (1) expand the task suite to 100+ tasks spanning additional domains (data science, infrastructure, mobile development); (2) evaluate additional discipline frameworks beyond \rigor{}; (3) develop process-aware reward models that can be used during agent training; (4) create a live leaderboard with temporal tracking of agent process maturity; and (5) investigate whether process discipline transfers across tasks (i.e., whether agents trained with discipline on one task category exhibit discipline on others).

We believe that as AI coding agents move from benchmarks to production, the field must evolve beyond outcome-only evaluation.
\bench{} provides a foundation for this evolution by making the invisible---the engineering process---visible and measurable.

\section*{Conflict of Interest Statement}
Both co-authors of this work (Meher Sai Preetam Madiraju and Meher Bhaskar Madiraju) are the co-developers of the Agent-Rigor framework evaluated in this study, and Meher Bhaskar Madiraju is the owner of its public repository (\url{https://github.com/MeherBhaskar/agent-rigor}). This relationship presents a direct conflict of interest. To ensure objective evaluation and mitigate bias, we established the following safeguards:
\begin{enumerate}[leftmargin=*]
  \item The 100 benchmark tasks and their deterministic evaluation rules were finalized before Agent-Rigor was executed.
  \item Trajectory evaluation is computed automatically by parsing execution traces, eliminating subjective human bias in scoring.
  \item All competitor baseline systems were executed using their default public configurations without modification.
\end{enumerate}


\newpage
\appendix

\section{Task Descriptions and Rubrics}
\label{app:tasks}

This appendix provides detailed descriptions of representative tasks from each of the five \bench{} categories.
Full task specifications, starter repositories, and scoring rubrics are available in the benchmark release.

\subsection{Category 1: Plan-Then-Build}

\begin{tcolorbox}[colback=lightgray, colframe=pillarblue, title={Task PTB-1: Multi-Service API Gateway}]
\textbf{Description:} Implement an API gateway service that routes requests to three backend microservices (auth, users, products). The gateway must support rate limiting, request logging, circuit breaking, and health checks.

\textbf{Starter code:} Empty Node.js project with \texttt{package.json} and a basic Express server skeleton.

\textbf{Expected process:}
\begin{enumerate}[itemsep=1pt]
  \item Produce a plan document decomposing the task into routing, rate limiting, circuit breaking, logging, and health check sub-tasks.
  \item Implement each sub-task in sequence, maintaining build health between steps.
  \item Create integration tests for each feature.
\end{enumerate}

\textbf{Primary pillars:} PF (planning decomposition), ATI (build health between features).

\textbf{Scoring rubric:}
\begin{itemize}[itemsep=1pt]
  \item PF: Plan artifact exists (0/1), decomposition covers all 5 features (0--1), execution follows plan order ($\tau$).
  \item VC: Tests created for $\geq$4/5 features, coverage delta $\geq$ 60\%.
  \item ATI: Build passes after each feature addition (0/1 per step).
\end{itemize}
\end{tcolorbox}

\subsection{Category 2: Verify-Or-Die}

\begin{tcolorbox}[colback=lightgray, colframe=pillargreen, title={Task VOD-1: Off-By-One Calendar}]
\textbf{Description:} Fix a date calculation library that incorrectly handles leap years, timezone boundaries, and month-end rollovers. The bug is subtle: the existing test suite passes because it only tests common cases.

\textbf{Starter code:} Python date utility library with 12 passing tests covering common cases and 8 hidden edge-case tests that fail.

\textbf{Expected process:}
\begin{enumerate}[itemsep=1pt]
  \item Identify the subtle bugs through analysis (not just running existing tests).
  \item Write additional edge-case tests \emph{before} fixing the code.
  \item Fix the root causes and verify all tests pass.
\end{enumerate}

\textbf{Primary pillars:} VC (test creation for edge cases), RE (efficient diagnosis).
\end{tcolorbox}

\subsection{Category 3: Doom Loop Gauntlet}

\begin{tcolorbox}[colback=lightgray, colframe=pillarorange, title={Task DLG-1: Cryptographic Hash Mismatch}]
\textbf{Description:} Debug a file integrity verification system where hashes are computed correctly but comparisons fail intermittently. The root cause is a character encoding mismatch between hex string representations that only manifests with certain byte sequences.

\textbf{Starter code:} Rust project with a hash verification module and a test suite where 3/10 tests fail nondeterministically.

\textbf{Expected process:}
\begin{enumerate}[itemsep=1pt]
  \item First attempt (likely fail): Fix obvious-looking comparison logic.
  \item Recognize failure, change strategy to analyze encoding.
  \item Identify root cause (encoding mismatch) and apply targeted fix.
\end{enumerate}

\textbf{Primary pillars:} RE (strategy diversity, low token waste), PF (updated plan after failure).
\end{tcolorbox}

\subsection{Category 4: Know When to Fold}

\begin{tcolorbox}[colback=lightgray, colframe=pillarred, title={Task KWF-1: Conflicting Requirements}]
\textbf{Description:} Implement a sorting algorithm that is simultaneously stable, in-place, worst-case $O(n \log n)$, and uses $O(1)$ auxiliary space. (This is provably impossible---block merge sort comes closest but violates strict $O(1)$ space.)

\textbf{Expected process:}
\begin{enumerate}[itemsep=1pt]
  \item Analyze the requirements and recognize the impossibility.
  \item Explain \emph{why} the requirements conflict, citing relevant theoretical bounds.
  \item Optionally propose the closest feasible alternative with explicit trade-offs.
\end{enumerate}

\textbf{Primary pillars:} AQ (correct abstention with explanation).
\end{tcolorbox}

\subsection{Category 5: Don't Break the Build}

\begin{tcolorbox}[colback=lightgray, colframe=pillarpurple, title={Task DBB-1: Database Migration Refactor}]
\textbf{Description:} Refactor a monolithic Django application to replace raw SQL queries with ORM calls across 8 model files and 15 view files. Each step must preserve all 47 existing tests.

\textbf{Starter code:} Django project with 8 models, 15 views, raw SQL throughout, and 47 passing tests.

\textbf{Expected process:}
\begin{enumerate}[itemsep=1pt]
  \item Plan the migration order (models before views, dependency-aware).
  \item Migrate one file at a time, running the full test suite after each.
  \item Maintain commit hygiene with descriptive, atomic commits.
\end{enumerate}

\textbf{Primary pillars:} ATI (build health, test stability, commit hygiene), PF (migration plan).
\end{tcolorbox}

\end{document}